\documentstyle[aps,prd]{revtex}

\def\D {\mbox{D}}
\def\div {\mbox{div}\,}
\def\rd {\displaystyle{\cdot}}
\def\c {\mbox{curl}\,}
\def\ep {\varepsilon}

\def \ts {\textstyle}
\def\be {\begin{equation}}
\def\ee {\end{equation}}
\def\bea {\begin{eqnarray}}
\def\eea {\end{eqnarray}}

\begin{document}

\title{Anisotropic stresses in inhomogeneous universes}

\author{John D. Barrow} 

\address{Astronomy Centre, University of Sussex, Brighton~BN1~9QJ,
UK}

\author{Roy Maartens}

\address{School of Computer Science and Mathematics, University of
Portsmouth, Portsmouth~PO1~2EG, UK}

\date{August 1998}

\maketitle

\begin{abstract}

Anisotropic stress contributions to the gravitational field
can arise from magnetic fields, collisionless relativistic
particles, hydrodynamic shear viscosity, gravitational waves,
skew axion fields in
low-energy string cosmologies, or topological defects.
We investigate the effects of such stresses on cosmological
evolution, and in particular on the dissipation of shear anisotropy.
We generalize some previous results that were given
for homogeneous anisotropic universes, 
by including small inhomogeneity in the
universe. This generalization is facilitated by a
covariant approach.
We find that anisotropic stress dominates the evolution of
shear, slowing its decay. The effect is strongest in 
radiation-dominated universes, where there is slow logarithmic
decay of shear.

\end{abstract}


\section{Introduction}

In this paper we 
consider the effects of anisotropic stresses on the late-time 
behavior of
anisotropic and inhomogeneous cosmological models which are close to
isotropy and homogeneity. We use an approximation scheme that is
similar to the velocity-dominated approximation but extend it to deal 
with
the presence of anisotropic stresses. Past investigations of the 
late-time
evolution of cosmological models in general relativity and in 
scalar-tensor
gravity theories have been performed under the assumption that the 
matter
source in the universe is composed only of isotropic perfect fluids.
However, from the study of homogeneous anisotropic universes, it is 
known
that the evolution of deviations from perfect isotropy is dominated 
by the
distortions created by any anisotropic stresses. Examples of such 
stresses
are provided by collisionless particles, gravitons, electric or 
magnetic
fields, or by topological defects, as discussed in \cite{b}.

In \cite{mes}, a covariant and
gauge-invariant approach was taken to the study of 
generic
distortions to the Friedmann universe in the dust dominated era. The
kinematic quantities which define deviations from perfect isotropy and
homogeneity were related to the multipoles of microwave background
temperature distribution. This enables limits to be placed on all 
forms of
anisotropic and inhomogeneous distortions of the spacetime metric 
without
making specific assumptions (like homogeneity) about the form of those
distortions \cite{mes2}.

In this paper we combine the approaches of \cite{b} 
and \cite{mes} in order to 
provide
a gauge-invariant and 
covariant description of the cosmological evolution 
of anisotropic
and inhomogeneous universes containing a perfect fluid and anisotropic
stress. 
This requires an extension of the covariant formalism
developed earlier in \cite{mes} to 
general anisotropic stresses and energy flux
terms, and to the radiation 
era. The covariant approach is based on the full nonlinear
equations, so that it naturally incorporates nonlinear
effects. By
the introduction of a self-consistent approximation scheme we 
determine the
time-evolution of the anisotropic stresses and the ratio of the 
densities of
the anisotropic to the isotropic fluid stresses. 

As in the case of 
spatially
homogeneous universes \cite{b,bs,b2}, the evolution of trace-free 
anisotropic
stresses during the radiation era represents particular mathematical
problems because of the nonlinear character of the anisotropic stress
perturbations. 
If linearized about the exact Friedmann expansion this 
case
would give rise to a zero eigenvalue. The dust evolution is determined 
in
the linear approximation about the Friedmann solution but the decay of 
the
shear anisotropy is considerably slowed by the presence of the 
anisotropic
stresses. In each case there is an attractor solution relating the
distortion in the expansion anisotropy to the fractional density in
anisotropic stress. 
Here we find that similar behavior persists in the presence of
small inhomogeneity.

\section{Covariant Lagrangian dynamics}

The source of the gravitational field is a 
non-interacting mixture of perfect and anisotropic fluids, 
the former typically representing isotropic radiation
or cold matter (dust),
while the latter encompasses an ultra-relativistic fluid,
collisionless or viscous, a magnetic field,
and other possibilities.
The total energy density is 
\[
\rho_{{\rm tot}}=\rho_{{\rm perf}}+\rho_{{\rm aniso}} 
\equiv \rho+\mu\,, 
\]
and the total isotropic pressure is 
\[
p_{{\rm tot}}=(\gamma-1)\rho+{\ts{1\over3}}\mu\,, 
\]
where $\gamma$ is constant, with $\gamma={4\over3}$ for radiation
and $\gamma=1$ for dust.
The perfect fluid is dynamically isotropic about a unique
4-velocity $u^a$, where $u_au^a=-1$. 
Relative to $u^a$,
the energy flux of the anisotropic fluid is $q^a$ 
and its anisotropic stress is $\pi_{ab}$, so that
the total energy--momentum tensor is 
\be
T_{ab}=(\rho+\mu)u_au_b+[(\gamma-1)\rho+{\ts{1\over3}}\mu]h_{ab}
+q_au_b+u_aq_b+\pi_{ab} \,,
\label{t}\ee
where $h_{ab}=g_{ab}+u_au_b$ is the projection tensor into the 
instantaneous rest spaces of comoving observers,
with $g_{ab}$ the spacetime metric. 

The anisotropic dynamic quantities $q_a$ and $\pi_{ab}$ can in 
general encode either relative velocity effects or dissipative 
effects, or both. The total energy flux in general comprises
a heat flux $Q_a$ and a particle diffusion flux (momentum
density) \cite{mt}
\[
q_a=Q_a+{\ts{4\over3}}\mu v_a \,,
\]
where $v_a$ is the velocity relative to $u_a$.
In the case of a magnetic field, $\pi_{ab}$ arises from
the anisotropic contribution of the field to pressure. For
a fluid in the hydrodynamic regime, $\pi_{ab}$
is driven by shear viscosity. For a collisionless gas,
$\pi_{ab}$ arises from the covariant quadrupole moment of the
distribution function.
The full anisotropy
information is encoded not only in $q_a$ and $\pi_{ab}$, 
but also in the
octopole and higher covariant 
multipoles of the distribution function \cite{mes}.

The covariant Lagrangian approach \cite{e} is based on a physical
4-velocity $u^a$, with all physical quantities represented
by tensors in the comoving instantaneous rest space. This
involves a
$1+3$ splitting via $u^a$ and 
$h_{ab}$. The covariant time 
and spatial derivatives are\footnote{
In \cite{mes}, $\D_a$ is denoted $\widehat{\nabla}_a$, and 
in \cite{eb} it is $^{(3)}\nabla_a$.
} 
$\dot{S}^{a\cdots}{}{}_{b\cdots}=u^c\nabla_c
S^{a\cdots}{}{}_{b\cdots}$ 
and $\D_cS^{a\cdots}{}{}_{b\cdots}=h_c{}^fh^a{}_d
\cdots h_b{}^e\cdots \nabla_fS^{d\cdots}{}{}_{e\cdots}$. 
(Note that $\D_ch_{ab}=0$.) 
As part of a streamlining and development of the covariant
formalism, we define
a covariant spatial divergence and curl of vectors and rank-2
tensors \cite{m}: 
\bea
&& \div V=\D^aV_a\,,~~~~(\div S)_a=\D^bS_{ab}\,, \\
&& \c V_a=\ep_{abc}\D^bV^c\,,~~~~ 
\c S_{ab}=\ep_{cd(a}\D^cS_{b)}{}^d \,,
\eea
where $\ep_{abc}=\eta_{abcd}u^d$ is the covariant spatial 
alternating tensor,
projected from the spacetime 
alternating tensor $\eta_{abcd}=-\sqrt{|g|}\delta^0{}_{[a}
\delta^1{}_b\delta^2{}_c\delta^3{}_{d]}$. (It satisfies 
$\ep_{abc}\ep^{def}=3!h_{[a}{}^dh_b{}^eh_{c]}{}^f$.)
These operators are irreducible parts of the spatial derivative,
making transparent the electromagnetic structure
of the general relativistic equations,
and they obey covariant identities 
which greatly facilitate manipulation of the equations to
investigate perturbations, integrability, wave equations, 
and so on (see \cite{m,mb}).

A projected vector satisfies 
$V_a=V_{\langle a\rangle}\equiv h_{ab}V^b$.
The covariant irreducible decomposition of a 
spatial rank-2 tensor is 
$S_{ab}={\ts{1\over3}}S_{cd}h^{cd}h_{ab} 
+S_{\langle ab\rangle }+\ep_{abc}S^c$,
where the PSTF part is $S_{\langle ab\rangle }=
h_{(a}{}^ch_{b)}{}^dS_{cd}- {\ts{1\over3}}S_{cd} h^{cd}h_{ab}$, 
and the skew
part is the dual of the spatial vector 
$S_a= {\ts{1\over2}}\ep_{abc}S^{[bc]}$.

The kinematic quantities arise from decomposing $\nabla_bu_a$: 
\bea
\nabla_bu_a &=& \D_bu_a-A_au_b \,,\\
\D_bu_a &=& {\ts{1\over3}}\Theta
h_{ab}+\sigma_{ab}+\ep_{abc} \omega^c \,,
\eea
where $A_a\equiv\dot{u}_a=A_{\langle a\rangle }$ 
is the acceleration, $\Theta=\D^au_a$ 
is the volume expansion rate, 
$\sigma_{ab}=\D_{\langle a}u_{b\rangle }$ is the shear, 
and $\omega_a=-{\ts{1\over2}}\c u_a$ is the vorticity. 
The Weyl tensor, representing the locally free
gravitational field \cite{mb}, is 
decomposed into gravito-electric and magnetic fields: 
\be
E_{ab}=C_{acbd}u^cu^d=E_{\langle ab\rangle }\,,
~~~~ H_{ab}={\ts{1\over2}}\ep_{acd}C^{cd}{}{}_{be}u^e
=H_{\langle ab\rangle} \,.
\ee
Then the background Friedmann model is covariantly 
characterized as the limiting case
\bea
\mbox{dynamics:}~~~ &&\D_a\rho=0=\D_a\mu\,, 
~q_a=0\,,~\pi_{ab}=0 \,, \label{frw1}\\
\mbox{kinematics:}~~~ && \D_a\Theta=0\,,~A_a=0=\omega_a\,,~
\sigma_{ab}=0 \,, \label{frw2}\\
\mbox{free field:}~~~ && E_{ab}=0=H_{ab} \,.\label{frw3}
\eea
In covariant and gauge-invariant perturbation theory \cite{eb}, 
the above
quantities are non-zero but small, so that the universe is almost 
isotropic and homogeneous.
In spatially homogeneous anisotropic models (Bianchi and
Kantowski-Sachs models), out of the quantities in
equations (\ref{frw1})--(\ref{frw3}), $q_a$, $\pi_{ab}$,
$\sigma_{ab}$, $E_{ab}$ and $H_{ab}$ may be nonzero. 

The field equations $R_{ab}=T_{ab}-{\frac{1}{2}}T_c{}^cg_{ab}$, 
with $T_{ab}$
given by Eq. (\ref{t}), are incorporated into the Ricci 
identity 
$\nabla_{[a}\nabla_{b]}u_c={1\over2}R_{abcd}u^d$ and
the Bianchi identities
$\nabla^dC_{abcd}=\nabla_{[a}(-R_{b]c}+{1\over6}Rg_{b]c})$, 
which are then covariantly decomposed into
propagation and constraint equations as in \cite{e}. 
The equations are significantly simplified and made clearer 
in the developed version of the formalism \cite{m}.
This
gives the following (exact, nonlinear) equations:\\

\noindent Propagation: 
\begin{eqnarray}
\dot\rho+\gamma \Theta\rho &=& 0\,,\label{c1}\\
\dot\mu+{\ts{4\over3}}\Theta\mu
+\D^aq_a &=& -\sigma^{ab}\pi_{ab}-2A^aq_a\,,
\label{c2}\\
\dot{q}_{\langle a\rangle}+{\ts{4\over3}}\Theta q_a+
{\ts{4\over3}}\mu A_a+{\ts{1\over3}}\D_a\mu+(\div\pi)_a &=&
-(\sigma_{ab}+\ep_{abc}\omega^c)q^b-A^b\pi_{ab}\,,\label{c3}\\
\dot{\Theta}+{\ts{1\over3}}\Theta^2 +
{\ts{1\over2}}\left[(3\gamma-2)\rho+2
\mu\right]-\D^aA_a &=&-\sigma_{ab}\sigma^{ab} 
+2\omega_a\omega^a+A_aA^a \,, \label{a1} \\
\dot{\omega}_{\langle a\rangle }+{\ts{2\over3}}\Theta\omega_a 
+{\ts{1\over2}}
\c A_a &=&\sigma_{ab}\omega^b  \,,\label{a2} \\
\dot{\sigma}_{\langle ab\rangle }+{\ts{2\over3}}\Theta\sigma_{ab} 
+E_{ab}-{\ts{1\over2}}\pi_{ab} -\D_{\langle a}A_{b\rangle } &=&
-\sigma_{c\langle a}\sigma_{b\rangle }{}^c- \omega_{\langle
a}\omega_{b\rangle } +A_{\langle a}A_{b\rangle }  
\,,\label{a3} \\
\dot{E}_{\langle ab\rangle }+\Theta E_{ab} -\c H_{ab} 
+{\ts{1\over2}} \dot{\pi}_{\langle ab\rangle }
+{\ts{1\over6}}\Theta\pi_{ab} &&  \nonumber \\
{}+{\ts{1\over2}}(\gamma\rho+{\ts{4\over3}}\mu)\sigma_{ab} 
+{\ts{1\over2}}\D_{\langle a}q_{b\rangle } &=& 
\sigma_{c\langle a}\left[3E_{b\rangle }{}^c -{\ts{1\over2}}
\pi_{b\rangle}{}^c\right]
-\omega^c \ep_{cd(a}\left[E_{b)}{}^d +{\ts{1\over2}}\pi_{b)}{}^d
\right]
\nonumber \\
&&{}-A_{\langle a}q_{b\rangle } +2A^c\ep_{cd(a}H_{b)}{}^d  
\,,\label{a4} \\
\dot{H}_{\langle ab\rangle }+\Theta H_{ab} +\c E_{ab} 
-{\ts{1\over2}}\c\pi_{ab} &=& 3\sigma_{c\langle a}H_{b\rangle }{}^c 
-\omega^c \ep_{cd(a}H_{b)}{}^d -2A^c\ep_{cd(a}E_{b)}{}^d  
\nonumber \\
&&{} -{\ts{3\over2}}\omega_{\langle a}q_{b\rangle }+ {\ts{1\over2}}
\sigma^c{}_{(a}\ep_{b)cd}q^d \,. \label{a5}
\end{eqnarray}\\

\noindent Constraint: 
\begin{eqnarray}
\gamma\rho A_a+(\gamma-1)\D_a\rho &=&0 \,,\label{c4}\\
{\ts{2\over3}}\D_a\Theta-(\div\sigma)_a+\c\omega_a -q_a&=& 
2\ep_{abc}\omega^bA^c\,,  \label{a6} \\
\div\omega &=&A^a\omega_a \,, \label{a7} \\
H_{ab}-\c\sigma_{ab}-\D_{\langle a} \omega_{b\rangle }&=&
2A_{\langle a} \omega_{b\rangle }\,,  \label{a8} \\
(\div E)_{a}-{\ts{1\over3}}\D_a(\rho+\mu) +{\ts{1\over2}}\D^b\pi_{ab}
+{\ts{1\over3}}\Theta q_a &=& \ep_{abc}\sigma^b{}_dH^{cd} 
-3H_{ab}\omega^b +{\ts{1\over2}}\sigma_{ab}q^b
+{\ts{3\over2}} \ep_{abc}\omega^bq^c \,, \label{a9}\\
(\div H)_{a}-(\gamma\rho+{\ts{4\over3}}\mu) \omega_a 
+{\ts{1\over2}}\c q_a &=&
-\ep_{abc}\sigma^b{}_d\left[E^{cd}+{\ts{1\over2}}\pi^{cd}\right]
+\left[3E_{ab} -{\ts{1\over2}}\pi_{ab}\right] \omega^b\,.  \label{a10}
\end{eqnarray}

We emphasise that these are the exact equations of the full
nonlinear theory, before any approximations are made.
They generalize 
the equations given for the case of irrotational dust in 
\cite{m} and perfect fluid in \cite{mb}, 
and the linearized dissipative
fluid equations given in \cite{mt}. Linearization about a
limiting background Friedmann model reduces all of the right hand 
sides to zero, in view of equations (\ref{frw1})--(\ref{frw3}).
(Linearization also allows us to remove the
angled brackets from the indices of time derivatives.)

We note that if $A_a=0$, then Eq. (\ref{c4}) implies $\D_a\rho=0$
unless $\gamma=1$. 
It then
follows from the energy conservation equation
(\ref{c1}) (recall that the two fluids are non-interacting)
and the identity 
\[
\D_a\dot{\phi}=\left(\D_a \phi\right)^{\rd}+{\ts{1\over3}}\Theta
\D_a\phi 
+\sigma_a{}^b\D_b\phi+\ep_{abc}\omega^b\D^c\phi-A_a\dot{\phi} 
-u_aA^b\D_b\phi \,,
\]
that $\D_a\Theta=0$ also. Furthermore, the identity
\[
\c\D_a\phi=-2\dot{\phi}\omega_a \,,
\]
shows that $\omega_a=0$. In summary: if the perfect fluid
is not dust, then vanishing acceleration ($A_a=0$) 
implies vanishing vorticity ($\omega_a=0$) and spatial
gradients of energy density and expansion rate 
($\D_a\rho=0=\D_a\Theta$).

The anisotropic stress 
$\pi_{ab}$ may be subject to an evolution equation,
i.e. it may be self-consistently determined 
via kinetic theory \cite{mes}
or thermodynamics \cite{mt,sb} or some other 
dynamics \cite{b,bl,tb}.
For example, in the case of a collisionless 
gas, the covariant quadrupole of the
Boltzmann equation
is a propagation equation for $\pi_{ab}$ \cite{sme}: 
\bea
&&\dot{\pi}_{\langle ab\rangle}+{\ts{4\over3}}\Theta\pi_{ab} 
+{\ts{8\over15}}\mu\sigma_{ab}+2\D_{\langle a}q_{b\rangle}
+(\div \xi)_{ab}= \nonumber\\
&&~-2A_{\langle a}q_{b\rangle}-{\ts{2\over7}} 
\sigma^c{}_{\langle a}\pi_{b\rangle c}+2\omega^d\ep_{cd(a} 
\pi_{b)}{}^c+\sigma^{cd}\chi_{abcd} \,,
\label{pi}\eea
where $\xi_{abc}$ and $\chi_{abcd}$ are the octopole and
hexadecapole. Another example is
a dissipative fluid, for which the causal thermodynamics of 
Israel and Stewart gives \cite{mt}
\be
\tau\dot{\pi}_{\langle ab\rangle}+\pi_{ab}=-2\eta\sigma_{ab}\,, 
\label{is}\ee
where $\eta$ is the shear viscosity and 
$\tau$ is the relaxation time-scale,
with $\tau=0$ in the non-causal Navier-Stokes-Fourier theory.
The heat flux in such a fluid obeys the causal transport
equation\footnote{
This neglects any coupling of heat flux and
viscosity, and assumes the relaxation time-scales for heat
conduction and shear viscosity are equal. For the general case,
see \cite{mt}.
}
\be
\tau\dot{Q}_{\langle a\rangle}+Q_a=-\kappa\left(\D_a T
+TA_a\right)
\,,
\label{is2}\ee
where $\kappa$ is the thermal conductivity and $T$ is the temperature.

In this paper, we will not impose evolution
equations such as (\ref{pi})--(\ref{is2}), since our main interest
is in the general dynamics of anisotropic stress, rather than
particular physical models of stress. For this purpose, we will
introduce in the next section a general expression for
$\pi_{ab}$.

\section{Approximations and Solutions}

The equations of the previous section provide in principle a
complete, fully covariant and nonlinear
description of the effects of anisotropic stresses
on cosmological evolution, purely in terms of quantities with
a direct and transparent physical interpretation.
However, the complexity of the exact nonlinear system
effectively rules out such a description, and we must resort
to approximations. The fundamental approximation is that the
universe is close to a Friedmann model. This in fact follows
after last scattering as a theorem \cite{sme}, using the observed
small anisotropy in the microwave background temperature, and
assuming a weak Copernican principle (which is falsifiable via
the Sunyaev-Zeldovich effect \cite{mes2}). Before last
scattering, in the radiation era, it is reasonable to make
the assumption because rapid collisions should tend to
damp out perturbations in the dominant component, leaving
anisotropic stresses of the decoupled component as a perturbation.

As pointed out above, it is not sufficient to linearize the
equations about the Friedmann limit, since during radiation
domination the anisotropic stress perturbations have an essential
nonlinear character, which is known in homogeneous anisotropic 
models \cite{bs,b2}, and will be confirmed to extend to the 
small-inhomogeneity case below.
A straightforward linearization would reduce all the right hand
sides of the propagation and constraint 
equations (\ref{c1})--(\ref{a10}) 
to zero. 
In order to incorporate the lowest order effects of anisotropic
stress and energy flux in the radiation-dominated era, we must retain
on the right hand sides those terms which couple $\pi_{ab}$ or
$q_a$ to the shear $\sigma_{ab}$.

Even when we neglect all second-order terms apart from those coupling
shear to anisotropic stress and energy flux, the equations
remain formidable. Since our main interest is in how the late-time
behavior of shear is affected by anisotropic stresses on large
scales, we impose the so-called velocity-dominated approximation,
i.e. that spatial derivatives are
neglected with respect to time derivatives. This means that in
the propagation equations (\ref{c1})--(\ref{a5}), we neglect spatial
derivative terms relative to the other terms.

We normalize the key quantities so as to
obtain a set of covariant dimensionless quantities. These are chosen
as follows:
\begin{eqnarray*}
&&\mu^* = {\mu\over\rho} \,,~\rho^*_a={\D_a\rho\over H\rho}\,,~ 
q^*_a={q_a\over\rho}\,,~ 
\pi^*_{ab}={\pi_{ab}\over\rho} \,,\\
&&\Theta^*_a={\D_a\Theta\over H^2}\,,~A^*_a={A_a\over H}\,,~
\omega^*_a={\omega_a\over H}\,,~\sigma^*_{ab}={\sigma_{ab}\over H}
\,,\\
&& E^*_{ab}={E_{ab}\over H^2}\,,~H^*_{ab}={H_{ab}\over H^2}\,,
\end{eqnarray*}
where $H=\dot{a}/a$ is the background Hubble rate. We also need the
background field equations
\[
\rho=3H^2(1+K)\,,~\dot{H}=-{\ts{1\over2}}H^2[3\gamma+(3\gamma-2)K]\,,
\]
where $K={\cal K}/(aH)^2$ and ${\cal K}=0,\pm 1$ is the curvature index.
The above dimensionless quantities are small, and vanish in the
background. Thus they are gauge-invariant as well as covariant
\cite{eb}.

Using these quantities, applying the velocity-dominated
approximation, and retaining terms that couple shear to anisotropic
stresses,
the propagation equations imply
\bea
\dot{\mu}^*&=&-(4-3\gamma)H\mu^* -H\sigma^{*ab}\pi^*_{ab}
\,, \label{np1}\\
\dot{q}^*_a &=&-(4-3\gamma)Hq^*_a -H\sigma^*_{ab}q^{*b}
\,, \label{np2}\\
\dot{\omega}^*_a &=& -{\ts{1\over2}}\left[4-3\gamma+(2-3\gamma)K
\right]H\omega^*_a
\,,\label{np3}\\
\dot{\sigma}^*_{ab} &=& -{\ts{1\over2}}\left[4-3\gamma+(2-3\gamma)K
\right]H\sigma^*_{ab}-HE^*_{ab}+{\ts{3\over2}}(1+K)H \pi^*_{ab}
\,, \label{np4}\\
\dot{E}^*_{ab} &=& -\left[3(1-\gamma)+(2-3\gamma)K\right]HE^*_{ab}
-{\ts{3\over2}}\gamma(1+K)H\sigma^*_{ab} 
-{\ts{3\over2}}(1+K)\dot{\pi}^*_{ab}\nonumber\\
&&~ +{\ts{3\over2}}(3\gamma-1)(1+K)H\pi^*_{ab}
-{\ts{3\over2}}(1+K)H\pi^*_{c\langle a}\sigma^*_{b\rangle}{}^c
 \,,\label{np5}\\
\dot{H}^*_{ab} &=& -\left[3(1-\gamma)+(2-3\gamma)K\right]HH^*_{ab}
+{\ts{3\over2}}(1+K)
H\sigma^*_{c(a}\ep_{b)}{}^{cd}q^*_d
\,. \label{np6}
\eea

Eq. (\ref{np2}) is a propagation equation for the
energy flux. There is no propagation equation
for the anisotropic stress, and we need a general
prescription which allows us to investigate qualitative
features of the role of anisotropic stress on shear
dissipation. Following previous studies in homogeneous
universes \cite{b,b2,old}, we assume that
the anisotropic stresses are of the form
\begin{equation}
\pi^*_{ab}=\mu^* \lambda _{ab}\,,~~\dot{\lambda}_{ab}=0\,,  \label{b}
\end{equation}
where $\lambda _{ab}$ is
a tensor whose components are slowly-varying functions of the space
coordinates $\vec{x}$
relative to the time variation of $\mu^*$.

We shall further assume that the universe 
is close to a flat Friedmann model, so that $K=0$.
The self-consistent
approximation for such an analysis is that the mean 
Hubble rate and the
perfect fluid density evolve, to leading order, as in the Friedmann
universe, so that
\begin{equation}
H=\frac 2{3\gamma t}\,,~~\rho =\frac 4{3\gamma ^2t^2}\,.  \label{c}
\end{equation}

Then the non-linearly coupled evolution equations (\ref{np1})
for $\mu^*$ 
and (\ref{np4}) for $\sigma^*_{ab}$
become
\begin{eqnarray}
\dot{\mu}^*+\left[\frac{2(4-3\gamma)}{3\gamma t}\right]\mu^* 
&=&-\left[\frac{2}{3\gamma t}\lambda
^{ab}\sigma^*_{ab}\right]\mu^*\,,  \label{f1} \\
\dot{\sigma}^*_{ab}+\left[\frac{(4-3\gamma )}{3\gamma t}\right]
\sigma^*_{ab}+\left(\frac{2}{3\gamma t}\right)
E^*_{ab} &=&\left(\frac{ \mu^*}{\gamma t}\right)
\lambda _{ab}\,.  \label{f2}
\end{eqnarray}

First, we note that a qualitatively different 
situation arises according as $\gamma
={4\over3}$ or $\gamma \neq {4\over3}$. 
If ${2\over3}<\gamma <{4\over3}$, then the right-hand side of
Eq. (\ref{f1}) can be neglected, and the evolution 
of $\mu^*$ is determined at linear
order in $\mu^*$ by the solution
\begin{equation}
\mu^*(t,\vec{x})=
M(\vec{x})t^{-n}~~~\mbox{where}~~~n={2(4-3\gamma)\over 3\gamma} \,,
\label{g}
\end{equation}
where $M$ is a slowly varying function of $\vec{x}$.
This expression also holds formally 
when $\gamma >{4\over3}$, but, since $n>0$ in
that case, the effects of the anisotropic 
stresses grow in time and the
assumption (\ref{c}) that 
the mean expansion is close to the Friedmann model, 
must eventually break down. Our analysis does not apply to this case.
When $0<\gamma \leq {2\over3}$, 
the anisotropic stresses do not dominate over the
simple $\sigma \propto t^{-2/\gamma }$ mode that 
arises when $\mu^*=0$. When 
${2\over3}<\gamma <{4\over3}$, 
$\mu^*$ decreases with time as $t\rightarrow \infty$, and the
approximation becomes increasingly better. However, the 
fall in $\sigma^*$ is
always at a slower rate than occurs in 
the absence of anisotropic stresses 
($\mu^*=0)$.
Note that when $\gamma =1$, we have 
$\mu^*\propto t^{-2/3}.$ 

However,
when $\gamma ={4\over3}$, the situation changes. 
The evolution of $\mu^*$ is then
determined by the nonlinear terms on the 
right-hand side of (\ref{f1}). In
this situation the universe evolves 
towards an attractor in which $\sigma^*_{ab}$ varies 
only logarithmically.

For general $\gamma \leq {4\over3}$, Eq. (\ref{f2})
has relaxation form, and we seek
asympotic solutions with $\sigma^*_{ab}\rightarrow0$. This gives
\begin{equation}
E^*_{ab}={\ts{3\over2}}\mu^*\lambda _{ab}+{\ts{1\over2}}
(3\gamma -4)\sigma^*_{ab}\,.  \label{i2}
\end{equation}
Hence for $\gamma<{4\over3}$
 \begin{equation}
E^*_{ab}\propto\sigma^*_{ab}\propto \mu^*\propto t^{-n}\,,
\label{i3}
\end{equation}
where the factors of 
proportionality are slowly-varying functions of the
space coordinates.

Specializing to the radiative case 
$\gamma ={4\over3}$ again, the evolution of the
gravito-electric field is determined by Eq. (\ref{np5}):
\begin{eqnarray}
\dot{E}^*_{ab} &=& \left({1\over 2t}\right)E^*_{ab}
-\left({1\over t}\right)
\sigma^*_{ab} 
+\left({9\over 4t}\right)\mu^*\lambda_{ab}
\nonumber\\
&&~ 
+\left({3\over 4t}\right)\mu^*\lambda^{cd}\sigma^*_{cd}
\lambda_{ab} 
-\left({3\over 4t}\right)\mu^*\lambda^c{}_{\langle a}
\sigma^*_{b\rangle c}
\,,\label{j}
\eea
Combining this with Eq. (\ref{i2}), 
neglecting all but leading order terms
in the shear, we have in this approximation 
\begin{equation}
\sigma^*_{ab}=3\mu^*\lambda _{ab}\,.  \label{k}
\end{equation}
Then, together with Eq. (\ref{f1}), this implies
\[
\dot{\mu}^*=-\frac {3}{2t}\mu^{*2}\lambda _{ab}\lambda ^{ab}\,,
\]
and therefore,
\begin{equation}
\mu^*(t,\vec{x})=\frac 2{3\lambda _{ab}(\vec{x})
\lambda ^{ab}(\vec{x})\left[N(\vec{x})+\ln t\right]}\,,  \label{m}
\end{equation}
where $N(\vec{x})$ is an arbitrary slowly-varying 
function of integration.
Thus, $\sigma^*_{ab}$ is determined in 
this approximation from Eq. (\ref{k}) and 
$E^*_{ab}$ is determined from Eq. (\ref{i2}). 
Asymptotically, as $t\rightarrow\infty $, we have
\begin{eqnarray}
&&E^*_{ab} \propto \pi^*_{ab}\propto 
\sigma^*_{ab} \propto  \mu^*\propto \frac 1{\ln t}\,,  \label{n2} \\
&& \sigma_{ab}\propto\frac 1{t\ln t}\,,~  
E_{ab}\propto \pi_{ab}\propto {1\over t^2\ln t} \,.\label{n3}
\end{eqnarray}
Moreover, in this approximation, using equations (\ref{i2}) 
and (\ref{k}), we have that
at large $t,$
\begin{equation}
E_{ab}={\ts{1\over2}}\pi _{ab}\,.  \label{o}
\end{equation}
Note that the shear evolution found here is dictated 
by the anisotropic
stresses. If we were to take $\pi _{ab}\equiv 0$, 
as is usually done in
analyses using the velocity-dominated approximation, 
then $\lambda _{ab}=0$
and we do not have $\sigma^*_{ab}\rightarrow (\ln t)^{-1}$ 
asymptotically.
Instead, we recover the familiar Bianchi type I mode with 
$\sigma_{ab}\propto a^{-3}\propto t^{-3/2},$ where 
$a$ is the geometric-mean
expansion scale factor ($H= \dot a/a$).

The propagation equation (\ref{np2})
for energy flux becomes, to
leading order,
\[
\dot{q}^*_a = -H\sigma^*_{ab}q^{*b}\,.
\]
Hence, in the $\gamma ={4\over3}$
case we have,
as $t\rightarrow \infty $, 
\[
\dot{q}^*_a=-\left[{1\over\left(\lambda _{cd}\lambda ^{cd}\right)
\,t\ln t}\right]\lambda_a{}^bq^*_b\,.
\]
It follows that
\be
q^*_a \propto {1\over\ln t}~\Leftrightarrow~q_a\propto 
{1\over t^2\ln t}\,,
\label{q}\ee
so that the energy flux has the same asymptotic behaviour as the
anisotropic stress.

\section{Conclusions}

In this paper we have set up an efficient covariant formalism for 
studying
the asymptotic behaviour of cosmological models. 
This formalism allows us to delineate the 
contributions
from linear and non-linear terms in the evolution. We have 
investigated the
influence of trace-free anisotropic stresses on the late-time 
evolution of
universes containing small anisotropies and inhomogeneities. 
These stresses
can be contributed by collisionless particles, magnetic fields, or
topological defects and are found to dominate the evolution of the
anisotropy. In the radiation-dominated case the contribution of 
trace-free
anisotropic stress tensors is found to be mathematically subtle. The
perturbations to isotropy appear at second-order in the shear and lead 
to a
slow logarithmic decay of the shear anisotropy. In the dust era the 
fall-off
is as a power-law in time but is again slower than it would be in the
absence of anisotropic stresses. These results generalize those found
earlier for homogeneous anisotropic models by Barrow 
\cite{b,b2} (see also \cite{old}).
They extend the analysis of Maartens et al \cite{mes} to the case 
where the background model model contains fluid with non-zero
pressure.
These analyses
show us that asymptotic studies of the late-time evolution of 
expanding
universes \cite{cdlp} 
need to include the effects of anisotropic stresses in order
to discover the leading-order form of the metric. 
This will also apply to
studies of the late-time evolution of cosmological models in other 
theories
of gravity, for example those in scalar-tensor gravity theories 
performed by
Cromer et al \cite{cdl}.

\[ \]
{\bf Acknowledgements}

JDB is supported by the PPARC.



\end{document}